# Thermodiffusion in model nanofluids by molecular dynamics simulations

G. Galliero[1,2,*], S. Volz[3]

[1]Laboratoire des Fluides Complexes (UMR-5150), Université de Pau et des Pays de l'Adour, BP 1155, F-64013 PAU Cedex, France.

[2]Laboratoire d'Etudes des Transferts d'Energie et de Matière, Université de Marne la Vallée-Paris Est, Bâtiment Lavoisier, Cité Descartes, Champs sur Marne, F-77454 Marne la Vallée Cedex 2, France.

[3]Laboratoire d'Energétique Moléculaire et Macroscopique, Combustion, Ecole Centrale Paris, Grande Voie des Vignes, F-92295 Chatenay-Malabry Cedex, France.

[*]Corresponding author: email: guillaume.galliero@univ-pau.fr;

Tel: +33 5 59 40 7704; Fax: +33 5 59 40 7695

**Abstract**

In this work, a new algorithm is proposed to compute single particle (infinite dilution) thermodiffusion using Non-Equilibrium Molecular Dynamics simulations through the estimation of the thermophoretic force that applies on a solute particle. This scheme is shown to provide consistent results for simple Lennard-Jones fluids and for model nanofluids (spherical non-metallic nanoparticles + Lennard-Jones fluid) where it appears that thermodiffusion amplitude, as well as thermal conductivity, decrease with nanoparticles concentration. Then, in nanofluids in the liquid state, by changing the nature of the nanoparticle (size, mass and internal stiffness) and of the solvent (quality and viscosity) various trends are exhibited. In all cases the single particle thermodiffusion is positive, i.e. the nanoparticle tends to migrate toward the cold area. The single particle thermal diffusion coefficient is shown to be independent of the size of the nanoparticle (diameter of 0.8 to 4 nm), whereas it increases with the quality of the solvent and is inversely proportional to the viscosity of the fluid. In addition, this coefficient is shown to be independent of the mass of the nanoparticle and to increase with the stiffness of the nanoparticle internal bonds. Besides, for these configurations, the mass diffusion coefficient behavior appears to be consistent with a Stokes-Einstein like law.

# 1. Introduction

Among the transport phenomena, thermodiffusion (Soret effect), which couples mass and heat fluxes[1], is probably the one for which the modeling of the corresponding transport property is the less satisfying in condensed phase despite recent improvements in simple fluids[2-4], polymers[5,6] or colloids[7-9]. It should be noted that this improvement is largely due to the fact that experimental measurements of this property, which are hard to handle because of thermal convection, have been strongly improved during the last decade and provide now consistent values using different experimental techniques[10].

One possibility to improve the modeling and the knowledge of the microscopic mechanisms of this cross effect is the use of molecular dynamics (MD) simulations on model systems. Such an approach has already proves its efficiency, at least concerning the microscopic mechanism responsible of thermodiffusion on rather academic systems, Lennard-Jones binary mixtures[11-14] and ternary ones[15], molecular fluids[16-19], associative mixtures[20], reactive mixtures[21], ionic systems[22], polymer[23], and fluids in porous media[24-26]. Concerning the modeling of thermodiffusion using MD results, we can mention in particular the work of Galliero et al.[27] on the mass effect in Lennard-Jones (LJ) mixtures and the work of Artola and Rousseau[28] which has shown the pertinence of some thermodynamic models to predict thermodiffusion in peculiar LJ mixtures.

Nevertheless, in dilute systems, thermodiffusion (or equivalently thermophoresis) was hardly accessible by molecular dynamics despite an obvious interest. This is due mainly because of the too poor statistics in such systems when using usual equilibrium or non-equilibrium MD algorithms[11,17,29]. In addition, when the solute particle is large compared to the solvent, as it is often the case in systems studied experimentally, the mass diffusion process becomes particularly low. So the simulation duration needed to perform the computation of thermodiffusion using MD becomes too large to be reasonably accessible despite the continuous increase in computational power.

Therefore, the main aim of this work is to provide a new MD procedure to allow the computation of infinite dilution (single particle) thermodiffusion in model systems. In particular, this new scheme will be applied to model nanofluids, i.e. dense fluids containing nanoparticles described as in Vladkov and Barrat[30], which have raised a large interest in thermal science due to their abnormally large thermal conductivity[31,32]. It should be nevertheless mentioned that the results provided here deal with small non-metallic nanoparticles (diameter up to 4nm) in a LJ fluid and not "true" nanofluids for which thermal conductivity has exhibited this special behavior.

The new MD scheme relies on the estimation of the thermophoretic force on a solute particle induced by a fluid subject to a thermal gradient. This enables, assuming a Stokesian behavior of the system, the direct knowledge of the single particle thermal diffusion factor (characteristic of the thermodiffusion amplitude). The approach proposed here is noted Single Particle Thermodiffusion Algorithm (SPTA) in the following. In addition, through the MD computation of the mass diffusion coefficient, the single particle thermal diffusion coefficient is deduced as well.

In a first part is provided the thermodiffusion formalism as well as the way the fluid and nanoparticles are modeled. In addition, the new molecular dynamics SPTA proposed is described in details together with the technical details of the simulations. Then, in a second part, the SPTA is validated through a comparison with what can be extrapolated from usual boundary driven Non-Equilibrium MD (NEMD) results for finite concentration, first on simple LJ mixtures and second on nanofluids. Finally, the SPTA is applied on various nanofluids for which the nature of the solvent (quality and viscosity) and of the solute nanoparticle (size, mass and internal stiffness) is changed to study the various influences of these physical parameters on the thermodiffusion in such model systems.

## 2. Theory and Modelling

### 2.1. Thermodiffusion

The diffusive mass flux of component 1, $\mathbf{J}_1$, in a binary mixture experiencing a temperature gradient and in mechanical equilibrium is well described, close to equilibrium, by[1]:

$$\mathbf{J}_1 = -\rho[D_{12}\nabla w_1 + w_1(1-w_1)D_T\nabla T] \quad (1)$$

where $\rho$ is the density, $w_1$ the mass fraction of component 1, $T$ the temperature, $D_{12}$ the mutual mass diffusion and $D_T$ the thermal diffusion coefficient (also termed thermophoretic mobility).

When the steady state is reached, the two contributions to the mass flux $\mathbf{J}_1$, due to mutual diffusion and to thermodiffusion, balance each other and the Soret coefficient can be defined as:

$$S_T = \frac{D_T}{D_{12}} = -\frac{1}{w_1(1-w_1)}\frac{\nabla w_1}{\nabla T} = -\frac{1}{x_1(1-x_1)}\frac{\nabla x_1}{\nabla T} \quad (2)$$

where $x_1$ is the molar fraction of component 1. To quantify the amplitude of the thermodiffusion effect relatively to that of mass diffusion it is more convenient to use the dimensionless thermal diffusion factor defined as :

$$\alpha_T = TS_T = T\frac{D_T}{D_{12}} \quad (3)$$

In this work, the solute is defined as component 1 and the solvent as component 2. Thus, a positive $\alpha_T$ means that the solute particles tend to migrate, relatively to the solvent, towards the cold areas.

### 2.2. Interactions potential

*2.2.1. Solvent*

To model the solvent-solvent interaction, the usual truncated Lennard-Jones 12-6 (LJ) potential is used on spheres:

$$U_{ij}^{LJ} = 4\varepsilon_{ij}\left[\left(\frac{\sigma}{r_{ij}}\right)^{12} - \left(\frac{\sigma}{r_{ij}}\right)^{6}\right] \quad (4)$$

where σ is the distance at which the potential is equal to zero (the "atomic diameter"), $\varepsilon_{ij}$ the potential depth and $r_{ij}$ the distance between atoms *i* and *j*. The cutoff radius, $r_c$, has been set to 2.5σ. LJ molecular parameters have been taken to be those of an Argon like fluid: σ=3.405 Å, ε=996 J.mol$^{-1}$. In addition, except when stated, a molar mass $m_s$=0.04 kg.mol$^{-1}$ has been used.

*2.2.2. Nanoparticles*

The nanoparticles are quasi-spherical and composed of atoms, having the same molecular parameters than those of the solvent, which are distributed on a FCC crystal as in the work of Vladkov and Barrat[30]. Interactions between nanoparticle atoms are described by a LJ potential, Eq. (4), and those within the nanoparticules are linked to their nearest neighbours through a Finite Extensible Nonlinear Elastic (FENE) bonding potential[33]:

$$U_{ij}^{FENE} = -0.5\kappa R_0^2 \ln\left[1 - \left(\frac{r_{ij}}{R_0}\right)^2\right] \quad (5)$$

where $R_0$ is a finite extensibility and $\kappa$ a spring constant. In this work, except when stated, the FENE parameters were set to $R_0$=1.5σ and $\kappa$=30ε/σ$^2$, which allows simulations with relatively large time steps. It should be noted that, with such parameters, the distance between bonded atoms is peaked on 0.97σ.

In addition, to avoid clustering effects, for atoms belonging to different nanoparticles, a cutoff radius equal to $2^{1/6}$σ (WCA potential[34]) has been taken which means that nanoparticles are purely repulsive to each other.

To modify the quality of the solvent (or equivalently the wetting properties between the nanoparticles and the solvent), the energy parameters of the cross interaction is expressed by:

$$\varepsilon_{ij} = k_{ij}\sqrt{\varepsilon_i \varepsilon_j} \qquad (6)$$

where $k_{ij}$ is a parameter taken between 0.5 (bad solvent) and 1.5 (good solvent) which is of great importance on the amplitude[14], and even the sign[28], of thermodiffusion in simple LJ fluids.

To estimate the diameter of a nanoparticle, as it presents some FCC facets, the radius of gyration, $R_g$, was estimated during the simulation using:

$$R_g^2 = \frac{\sum_{i=1}^{N_p}(r_i - r_{c.m.})^2}{N_p} \qquad (7)$$

where $N_p$ is the number of atoms in the nanoparticle and $r_{c.m.}$ the position of the nanoparticle centre of mass. Assuming a homogeneous mass distribution, the effective outer diameter of the nanoparticle, $d_{NP}$, can be estimated using:

$$d_{NP} = \left(\frac{20 R_g^2}{3}\right)^{1/2} \qquad (8)$$

In this work, various nanoparticle sizes have been simulated, see table I. In real units, as σ=3.405 Å, the range of nanoparticle diameter tested goes from 0.8 to 4 nm which remains small compared to usual nanofluids. In fact, it should be noted that a more usual diameter of 10 nm (which remains nevertheless a small value), would imply around $2.5 \; 10^4$ atoms in the nanoparticle ($d_{NP} \propto N_p^{1/3}$) and at least $2.5 \; 10^5$ in the fluid. This would lead to an inaccessible CPU time cost of at least $10^5$ hours on a recent supercomputer to obtain one value of the thermal diffusion factor by MD simulation.

### 2.3. Measurement of mass diffusion

In order to deduce the thermal diffusion coefficient from the thermal diffusion factor, one needs the mass diffusion coefficient, see Eq. (3). This coefficient has been obtained during separate equilibrium MD simulations from the mean-squared displacement of the centre of the mass of all fluid molecules, $\Delta r_i^{c.m.}$, using:

$$D_{12} = x_1 x_2 \left( \frac{1}{x_1 m_1} + \frac{1}{x_2 m_2} \right)^2 (x_2 m_2)^2 \frac{N \left( \left\langle \Delta r_i^{c.m.} \right\rangle \right)^2}{6t} \quad (9)$$

where $N$ is the total number of particles, $t$ the time and $m_i$ the mass of particle $i$.

## 2.4. Computation of the thermodiffusion: usual Non Equilibrium MD

To compute thermodiffusion in concentrated systems the Non Equilibrium Molecular Dynamics approach of Müller-Plathe and Reith[29] has been used. In this scheme, a biperiodical heat flux is imposed to the simulation, and after a transient state, the thermal diffusion factor can be deduced using eqs. (2-3) and the measured temperature and molar fraction profiles. An example, on a half simulation box, of the profiles obtained is shown on figure 1. It is interesting to note that, as expected from Eq. (2), the molar fraction gradient can be non linear in nanofluids when the nanoparticles molar fraction is low and $\alpha_T$ is large as in figure 1. In addition, as a by-product, the effective thermal conductivity can be deduced using the Fourier's Law and the temperature profile.

To impose the heat flux, the simulation box is divided into $N_s$ slabs, of identical thickness and volume. Slabs 1 and $N_s$ are defined as the "hot" ones and slabs $N_s/2$ and $N_s/2+1$ as the "cool" ones. The heat flux is generated by an exchange, at a given swapping frequency for both half simulation boxes, of the solvent particle that has the lowest kinetic energy (from the "hot" slab) with the one that has the highest kinetic energy (from the "cold" slab). The four slabs where the exchanges are performed, as well as their first neighbors, have been discarded to measure the profiles.

## 2.5. New scheme to compute the single particle thermal diffusion factor.

When the concentration of the solute becomes too small, the usual MD approaches to estimate thermodiffusion, such as the one presented in the previous section, are inefficient mainly because of the too poor statistics. In addition, when the solute particle is large compare to the solvent molecule, the mass diffusion process (associated to the thermodiffusion characteristic time[1]) is very slow and the MD simulations need to be performed for a very long duration. Therefore, in order to study the thermodiffusion in dilute nanofluids using molecular dynamics a new approach is proposed to deduce the single particle (SP) thermal diffusion factor, $\alpha_T^{SP}$, which corresponds to the infinite dilution limit.

Let us consider a single particle (solute) in a fluid (solvent) subjected to an established thermal gradient. This particle experiences a thermophoretic force, $\mathbf{F}_T$, which in turns will induce a velocity drift of the particle, $\mathbf{v}_T$, proportional to the temperature gradient in dilute systems:

$$\mathbf{v}_T = -D_T^{SP} \nabla T \qquad (10)$$

where $D_T^{SP}$ is the single particle thermal diffusion coefficient.

If we consider that this velocity drift is small enough (the particle Reynolds number is small in the cases treated here) and that the time of the analysis is long enough (typically on the order of a few nanoseconds), inertial forces can be neglected. In that case, the hydrodynamic force related to friction (Stokes drag) acting on the particle, $\mathbf{F}_h$, is linearly related to the velocity of the sphere, $\mathbf{v}$, through:

$$\mathbf{F}_h = -\xi \mathbf{v} \qquad (11)$$

where $\xi$ is the friction coefficient (usually expressed as proportional to the solvent viscosity and the hydrodynamic radius of the sphere[35]).

The force balance implies $\mathbf{F}_h + \mathbf{F}_T = 0$ which induces that the velocity drift due to the thermophoretic force writes as:

$$\mathbf{v}_T = \frac{\mathbf{F}_T}{\xi} \qquad (12)$$

Moreover, if we assume that a Stokes-Einstein law[36] applies between the single particle mass diffusion, $D_{12}^{SP}$, and friction, we have:

$$D_{12}^{SP} = \frac{k_B T}{\xi} \quad (13)$$

Finally combining Eq.(10) with Eqs.(12-13), we obtain that the single particle thermal diffusion factor, $\alpha_T^{SP}$, in such systems can be expressed as:

$$\alpha_T^{SP} = \frac{T D_T^{SP}}{D_{12}^{SP}} = -\frac{\mathbf{F}_T}{k_B \nabla T} \quad (14)$$

Such an expression implies that the measurement of the thermophoretic force acting on the particle for a given thermal gradient provides a straightforward estimation of the thermal diffusion amplitude in such dilute systems. In addition, if the single particle mutual diffusion coefficient, $D_{12}^{SP}$, is known, the single particle thermal diffusion coefficient can be deduced from:

$$D_T^{SP} = -\frac{D_{12}^{SP} \mathbf{F}_T}{k_B T \nabla T} \quad (15)$$

In order to measure the thermophoretic force, $\mathbf{F}_T$, acting on the particle using molecular dynamics simulations, a simple scheme is proposed.

First, an initial system composed of two particles and the fluid is constructed. The two particles that are centered around $L_x/4$ and $3L_x/4$, $L_x$ being the size along $x$ of the simulation box, are attached to the reference frame of the simulation box through a harmonic potential with a spring constant:

$$\kappa_h = 30 C \frac{\varepsilon}{\sigma^2} \quad (16)$$

where $C$ is a numerical constant below unity and set to ≈0.1 in order to weakly perturb the system. Then, the NEMD approach described in the previous section is applied to generate a bi-periodical linear thermal gradient to the simulation box in the direction $x$. After a transient state, the location of the centres of mass of the two particles will be displaced relatively to their point of fixation because the thermophoretic force applies on them. The measure of this displacement for both particles, $\Delta x$, provides an estimation of the amplitude of the thermophoretic force, $F_T$, that applies on each particle through:

$$F_T = \kappa_h \Delta x \tag{17}$$

To be consistent with the sign convention of the thermal diffusion factor, $F_T$ is counted positively when the displacement goes against the thermal gradient. It should be noted that this scheme is similar to the experimental approach used in Albanese et al.[37] to study thermal radiation forces produced by heat flow through solid slabs in nonisothermal liquids.

## 2.6. Technical details of the simulations

The simulation box is cubic. A reduced timestep, $\delta t^* = \delta t \varepsilon^{1/2} m^{-1/2} \sigma^{-1}$, varying from 0.001 to 0.004 has been used. To integrate the equation of motion the Velocity Verlet algorithm has been applied. In a first step, to obtain the desired temperature and pressure, Berendsen thermostat and barostat were used[38] during at least $10^5$ timesteps. Then, nonequilibrium molecular dynamics runs have been performed to collect the results. To avoid nonlinearity in the temperature profile and phase transitions during the NEMD simulations, a weak reduced energy flux, e.g. in figure 1, $J_U^* = J_U \sigma^3 \varepsilon^{-3/2} m^{1/2}$, ranging from 0.02 to 0.006, has been applied (which corresponds to an exchange period of 50-150 timesteps).

Simulations on nanofluids have been performed on systems composed of 4000 to 21794 atoms (solute +solvent) depending on the size of the nanoparticle to avoid as much as possible finite size effects and interactions between nanoparticles in the case of the new scheme proposed. All nanofluids simulations have been performed at $T^*=k_bT/\varepsilon=1$ and $P^*=P\sigma^3/\varepsilon=1$ which corresponds to a dense liquid state ($\rho^*=N\sigma^3/V \approx 0.8$ in a pure LJ fluid for such conditions). After the transient state, runs of 2 to 4 $10^7$ timesteps were used to obtain the thermal diffusion factor for classical NEMD simulations. Concerning the SPTA, four independent runs of 5-20 $10^6$ each have been carried out.

To generate the initial nanofluid systems a FCC bulk arrangement of atoms is constructed. Then, random atoms are chosen as centre of nanoparticles and, avoiding the overlap, the atoms located within the radius of the nanoparticles are linked to their first neighbours by the FENE bond.

Concerning simulations on simple fluids mixtures, as previously done[14], they have been performed on systems composed of 1500 LJ particles for duration varying from 2-8 $10^7$ nonequilibrium timesteps.

For the computation of mutual diffusion, four independent runs of $10^6$ timesteps have been performed at equilibrium for each system. As expressed in Eq. (9), the slope of the mean-squared displacement versus time was calculated during the linear regime.

## 3. Results

### 3.1. Preliminary study

#### *3.1.1. Simple fluids*

The first point was to analyse the validity of the proposed method to evaluate the single particle thermal diffusion factor, $\alpha_T^{SP}$, in simple dense LJ binary mixtures. In fact, this approach is based on "macroscopic" liquid state assumptions and is therefore, a priori, limited to large particle immersed in a liquid. Nevertheless, it has been shown that the definition of a Stokes-Einstein like law at this scale makes sense[39-40] with an appropriate friction factor, $\xi$. In addition, the final relation of Eq. (14) derived for $\alpha_T^{SP}$ only implies the thermal gradient and the thermophoretic force and so does not need an explicit formulation of the friction factor $\xi$.

The first mixture tested is composed of species having the same molecular parameters ($\sigma_1=\sigma_2$ and $\varepsilon_1=\varepsilon_2$) except that $m_1/m_2=10$ (isotope like mixtures) i.e. ideal ones in the thermodynamic sense. Two thermodynamic conditions have been studied, one corresponding to a dense liquid at $T^*=1$ and $\rho^*=0.8$ and the second one to a dense supercritical gas at $T^*=1.686$ and $\rho^*=0.477$ that was studied previously[14]. In addition, results have been compared with the correlation recently proposed[27] to predict thermodiffusion in such systems.

Results shown on figure 1 clearly reveal that, the proposed approach to compute the infinite dilution values of $\alpha_T$ is consistent with what could be extrapolated from the values obtained by classical NEMD simulations. It is interesting to note that, despite the underlying assumptions see section 2.5., this new approach is able to provide reasonable result for both molar fraction limits ($x_1 \rightarrow 0$ and $x_1 \rightarrow 1$) and for both states. Besides, the results provided by the correlation provided in Ref. 27 are in agreement with the MD ones despite non-negligible deviations (up to 15 %).

The second mixture is a $n$-decane/methane (modelled by LJ spheres) one at $T^*=1.686$ and $\rho^*=0.477$ previously studied[15]. By using the SPTA, it has been found that $\alpha_{T,x_1 \rightarrow 0}^{SP} = 8.7 \pm 1.9$ and $\alpha_{T,x_1 \rightarrow 1}^{SP} = 0.8 \pm 0.3$ where the extrapolation of the results obtained by "classical" NEMD yields $\alpha_{T,x_1 \rightarrow 0} = 9.55$ and $\alpha_{T,x_1 \rightarrow 1} = 0.935$ [15]. Thus, on the LJ mixtures studied (ideal or not, liquid or supercritical), the SPTA is able to provide a consistent estimation of $\alpha_T^{SP}$.

Besides, it is worth noting that the possibility of obtaining the $\alpha_T^{SP}$ coefficient for such fluid mixtures, for both infinite dilution limits, could be valuable as well because recently proposed models to estimate thermodiffusion[41-42] uses these data to predict the thermodiffusion on the whole molar fraction range.

*3.1.2. Nanofluids*

To test the validity of the SPTA when applied on nanofluids, which are the main concern of this work, nanofluids have been simulated for various nanoparticles volume fraction and for the two smallest sizes studied in this work, $d_{NP}/\sigma=2.4$ and 4.03 at $T^*=1$ and $P^*=1$ (approximately $T=120$ K and $P=42$ MPa for the molecular parameters chosen). For such conditions, we have computed that the solvent viscosity $\eta_s=1.8\pm0.1\ 10^{-4}$ Pa.s and the solvent self diffusion $D_s=3.6\pm0.1\ 10^{-9}$ m$^2$.s.

Results provided on Fig. 3 confirm that the proposed SPTA is able to provide an estimation of $\alpha_T^{SP}$ which is consistent with what could be extrapolated from classical NEMD results. This is true even it seems that the SPTA values slightly underestimate extrapolated values as in the previous section. This may come from the fact that for very low concentration $\alpha_T$ becomes independent of concentration as particles do not see each other as noticed experimentally by Ning et al.[43] or because of a weakness in the underlying theory as claimed by Brinquier and Bourdon[44].

Besides, whatever the size and the volume fraction, the nanoparticles tend to migrate towards the cold area; $\alpha_T$ being always positive and rather important as seen in Fig. 3. In addition, $\alpha_T$ decreases strongly with the nanoparticles volume fraction. More precisely, it seems that the values tend towards zero for large nanoparticles concentrations, i.e. >30% in volume. It was not possible to see whether or not a sign inversion may occur for larger volume fraction because such concentrations were not accessible by the scheme adopted to construct initial systems. The noted $\alpha_T$ evolution with respect to the nanoparticles concentration is consistent with the experiments and model proposed by Ning et al.[43] to describe the hard-sphere suspensions and the work of Rauch et al.[5] on polymer suspensions. Nevertheless the decrease noted here is larger than what can be expected from both models.

As a by-product of the NEMD simulations, the thermal conductivity can be obtained for the nanofluids studied. Figure 4 clearly shows that thermal conductivity tends to decrease with the nanoparticles volume fraction, this decrease being slightly more pronounced for the smallest nanoparticle. These trends can be explained by two antagonist effects, the thermal shortcuts induced by the presence of nanoparticles that are more conductive than the fluid[45] (volume effect), and the Kapitza resistance occurring at the interface nanoparticle-fluid[30] (surface effect). For the studied systems, the surface effects predominate over the volume ones.

It should be noted that the decrease of the thermal conductivity λ with the nanoparticles volume fraction is in contradiction with what is generally found in experimental reports on nanofluids[31-32] whereas most of MD simulations confirm our trend[30,46]. Nevertheless, the here studied systems are far from being similar to those that are experimentally analysed. In particular the nanoparticles are not metallic ones (and are hence less conductive), the nanoparticles are smaller than in experiments and the fluid is by far simpler than water. For instance, it does not involve association.

Besides, it is worth noticing that the thermal conductivity measured here is an effective one. Thus, it takes into account the transport of energy through the Dufour effect (the symmetric effect to the Soret one). Therefore, if $D_T$ is positive, this effect tends to slightly decrease the effective thermal conductivity[1] when the concentration gradient is established, but no more than a few percents.

**3.2. Single particle thermodiffusion**

In this section all simulations using the SPTA have been performed so that $T^*=1$ and $P^*=1$ in the fluid.

*3.2.1. Influence of the size*

Among the parameters that may affect the single particle thermodiffusion is the size (i.e the radius) of the nanoparticle, as already noticed in section 3.1.2. Simulations have been performed for $d_{NP}/\sigma$ ranging from 2.4 to 11.59 as reported in Table I.

From the values provided in the appendix, it first appears that the sign of $\alpha_T^{SP}$ is always positive whatever the size (i.e. nanoparticle tends to migrate towards the cold areas) and the obtained values are from one to two orders larger than the one for non interacting particles[8] ($\alpha_T$=1). In addition, $\alpha_T^{SP}$ amplitude increases with the diameter in a monotonic way and roughly linearly with $d_{NP}$.

If $D_T^{SP}$ is deduced from the computation of $D_{12}^{SP}$ and the results obtained on $\alpha_T^{SP}$, further information can be gathered as provided in Fig. 5. The most striking result is that $D_T^{SP}$ seems to be independent of the size of the nanoparticle, with a value that is always close to 0.7 $10^{-10}$ $m^2.s^{-1}.K^{-1}$. It should be noted that such behaviour is consistent with findings on dilute polymer[47-48] and nanodroplets[49]. Besides, as shown by figure 5, $D_{12}^{SP}$ is roughly proportional to the inverse of the radius of the nanoparticle which is consistent with a Stokes-Einstein law, Eq. (13), as ξ is proportional to the hydrodynamic radius.

### *3.2.2. Influence of the nature of the solvent*

In this section and the following, we have focused our work on a nanoparticle composed of 55 atoms ($d_{NP}/\sigma$=4.03).

To quantify the effect of the nature of the solvent on the single particle thermodiffusion, we have first evaluated the influence of the quality of the solvent by varying $k_{ij}$ from 0.5, which corresponds to a bad solvent, to 1.5, i.e. a good solvent, appearing in Eq. (7). In a second step, we have varied the viscosity of the solvent, without modifying its affinity with the solute, through the modification of $m_s$ from $m_s/m_{NP}$ =1 to $10^{-2}$, as solvent viscosity scales[50] as $m_s^{1/2}$.

Figure 6 shows that the quality of the solvent affects both thermodiffusion and mass diffusion. $D_T^{SP}$ tends to increase with the quality of the solvent (as $\alpha_T^{SP}$, see appendix). This is in agreement with previous MD works on different systems[23,28] and with a lattice model described in Ref. 51. But, in the range of the $k_{ij}$ values tested, no sign inversion has been noticed.

$D_{12}^{SP}$ has an opposite behaviour as proved by figure 6. It decreases, nearly linearly, with the quality of the solvent. Such behaviour can be understood by the fact that increasing $k_{ij}$ will imply that more solvent particle will be structured or adsorbed around the nanoparticle. As a consequence the hydrodynamic radius will be larger and the mass diffusion coefficient will decrease in consequence.

As clearly shown on Fig. 7, both $D_T^{SP}$ and $D_{12}^{SP}$ strongly decrease with the viscosity of the solvent. More precisely they are both nearly proportional to $\eta_s^{-1}$ (or equivalently to $D_s$). It should be noted that, concerning $D_T^{SP}$ in dilute polymer solutions, such dependence to the viscosity with a $\eta_s^{-1}$ behaviour, has been emphasized in a recent experimental/theoretical work[48].

These behaviours are consistent with the fact that increasing the solvent viscosity leads to an increase of the friction coefficient ξ that is usually assumed[35] to be proportional to $\eta_s$. Hence, this increase leads to a decrease of $D_{12}^{SP}$ and $D_T^{SP}$, Eqs. (13,15), if we assume that the thermophoretic force, $F_T$, is weakly dependent on $m_s$. Concerning this last point, the fact that $\alpha_T^{SP}$ slightly decreases with $m_s$, as shown in the appendix, implies that $F_T$ is not perfectly independent of $m_s$ for the system studied. This may be explained by the fact changing $m_s$, apart from modifying $\eta_s$, does affect the thermal conductivity of the solvent ($\lambda_s$ is proportional[50] to $m_s^{-1/2}$). Hence, we can suspect that $F_T$ is affected by a modification of the difference (or the ratio) between $\lambda_s$ and $\lambda_{NP}$, as explicitly done in some theoretical models to predict thermodiffusion[8,52].

### 3.2.3. Influence of the nature of the particle

To analyse a possible influence of the internal dynamic of the nanoparticle, two parameters were varied while keeping those of the solvent constant: the mass of the nanoparticle atoms, $m_{NP}$ (from $m_{NP}/m_s$=1 to 50), and the strength of the intramolecular interaction through the modification of the "spring" constant $\kappa$ (ranging from $10\varepsilon/\sigma^2$ to $90\varepsilon/\sigma^2$) appearing in the FENE potential, Eq. (5).

Contrary to what occurs in simple LJ fluids[27], results shown on Fig. 8 indicate that $D_T^{SP}$ seems to be independent of $m_{NP}$, whereas $D_{12}^{SP}$ decreases weakly with $m_{NP}$. This last result indicates that the particles are too small to behave like Brownian ones (for which $D_{12}^{SP}$ is independent of $m_{NP}$) as noted as well in previous works for similar systems[40,53].

Another interesting point is that by changing the mass $m_{NP}$, $\lambda_{NP}$ is modified because, as the FENE potential is weakly anharmonic, $\lambda_{NP}$ should roughly be proportional to $(\kappa/m_{NP})^{1/2}$. Therefore the ratio between $\lambda_{NP}$ and $\lambda_s$ evolves in the same way than in the previous section, i.e. proportionally to $(m_{NP}/m_s)^{-1/2}$. This may explain why $\alpha_T^{SP}$ evolves versus $m_{NP}/m_s$ in the same way than in the previous section, as shown in the appendix.

As shown in Fig. 9 and contrary to the mass $m_{NP}$, the nanoparticle internal stiffness, $\kappa$, which affects[45] $\lambda_{NP}$, has a non negligible influence on $D_T^{SP}$. The larger the spring constant $\kappa$ is, the larger $D_T^{SP}$ (and $\lambda_{NP}$). In addition $\alpha_T^{SP}$ increases with $\kappa$, see Appendix. This result clearly shows that the internal degrees of freedom of the nanoparticle affects thermodiffusion whatever is the coefficient to quantify it, $D_T^{SP}$ or $\alpha_T^{SP}$.

Concerning $D_{12}^{SP}$, its behaviour with $\kappa$ is similar to the one of $D_T^{SP}$ in a less pronounced manner as illustrated by Fig. 9. In fact, increasing $\kappa$ generates two effects: the nanoparticle is more conductive and the radius of the nanoparticle slightly decreases, e.g. $d_{NP}/\sigma=4.26$ when $\kappa=10\varepsilon/\sigma^2$ and $d_{NP}/\sigma=3.79$ when $\kappa=90\varepsilon/\sigma^2$. Thus, the hydrodynamic radius decreases with $\kappa$ and so $D_{12}^{SP}$ increases with $\kappa$ when assuming a Stokes-Einstein like behaviour.

## 4. Summary and Conclusions

In this work, we provide a new simple scheme to compute, using Non Equilibrium Molecular Dynamics simulations, the Single Particle (infinite dilution) thermodiffusion in dense fluids which is not accessible in a reasonable amount of time using standard MD scheme.

In a first part, by a comparison with extrapolations from finite concentration results using standard NEMD simulations, this scheme is shown to be efficient in ideal and non ideal Lennard-Jones binary mixtures as well as in model nanofluids. These last being described by a LJ fluid + quasi-spherical nanoparticles composed of atoms distributed on a FCC crystal and interacting through LJ potential plus FENE bonding with the nearest neighbours. For these systems it appears that the thermodiffusion and thermal conductivity amplitudes decrease with nanoparticles concentration for both nanoparticles sizes tested.

Then, using the proposed scheme, the influence of the nature of the solvent and of the particle on the single particle thermodiffusion is estimated in these model nanofluids. It appears that, in all cases studied here, the nanoparticle tends to migrate towards the cold area and that both the nature of the solvent and the nature of the particle affect the thermodiffusion amplitude. The most striking result is that the thermal diffusion coefficient is shown to be independent of the size of the nanoparticle on the range tested ($d_{NP}$ = 0.8-4 nm), whereas the mass diffusion coefficient is roughly proportional to the inverse of the nanoparticle diameter (consistent with the Stokes-Einstein law). Concerning the influence of the nature of the solvent, it appears that $D_T^{SP}$ (and $D_{12}^{SP}$) is nearly proportional to the inverse of the solvent viscosity and increases with the quality of the solvent. Besides, it is shown that $D_T^{SP}$ is unaffected by the mass of the nanoparticle atoms whereas $D_{12}^{SP}$ slightly decrease with $m_{NP}$ (i.e. nanoparticles studied here are not Brownian particles). Finally, results indicate that both $D_T^{SP}$ and $D_{12}^{SP}$ increase with the value of the internal stiffness of the nanoparticle (through the modification of the spring constant $\kappa$).

**Acknowledgments**

We thank the TREFLE laboratory in Bordeaux and the CINES in Montpellier for having provided a large part of the computer time needed for this work. F. Montel is acknowledged for fruitful discussions.

**Appendix :**

Single particle thermal diffusion factor results for the various nanofluids studied in this work.

| $d_{NP}/\sigma$ | $m_{NP}$ (kg.mol$^{-1}$) | $m_s$ (kg.mol$^{-1}$) | $\kappa\sigma^2/\varepsilon$ | $k_{ij}$ | $\alpha_T^{SP}$ |
|---|---|---|---|---|---|
| 2.4 | 0.04 | 0.04 | 30 | 1 | 8.4±1.5 |
| 4.03 | 0.04 | 0.04 | 30 | 1 | 20.1±4.1 |
| 5.98 | 0.04 | 0.04 | 30 | 1 | 35.4±7.4 |
| 7.73 | 0.04 | 0.04 | 30 | 1 | 52.8±10.5 |
| 9.75 | 0.04 | 0.04 | 30 | 1 | 62.5±15 |
| 11.59 | 0.04 | 0.04 | 30 | 1 | 91±50 |
| 4.03 | 0.04 | 0.02 | 30 | 1 | 19.8±3.9 |
| 4.03 | 0.04 | 0.008 | 30 | 1 | 24.0±6.4 |
| 4.03 | 0.04 | 0.004 | 30 | 1 | 23.5±7.6 |
| 4.03 | 0.04 | 0.002 | 30 | 1 | 26.1±7.8 |
| 4.03 | 0.04 | 0.0008 | 30 | 1 | 27.6±9.1 |
| 4.03 | 0.08 | 0.04 | 30 | 1 | 22.2±4.3 |
| 4.03 | 0.2 | 0.04 | 30 | 1 | 23.5±4.5 |
| 4.03 | 0.4 | 0.04 | 30 | 1 | 23.4±6.8 |
| 4.03 | 0.8 | 0.04 | 30 | 1 | 25.5±6.5 |
| 4.03 | 2.0 | 0.04 | 30 | 1 | 28.1±6.5 |
| 4.03 | 0.04 | 0.04 | 10 | 1 | 13.1±2.5 |
| 4.03 | 0.04 | 0.04 | 20 | 1 | 18.5±4.5 |
| 4.03 | 0.04 | 0.04 | 60 | 1 | 23.75±4.2 |
| 4.03 | 0.04 | 0.04 | 90 | 1 | 25.6±6.3 |
| 4.03 | 0.04 | 0.04 | 30 | 0.5 | 8.5±2.9 |
| 4.03 | 0.04 | 0.04 | 30 | 0.75 | 16.7±4.5 |
| 4.03 | 0.04 | 0.04 | 30 | 1.25 | 23.4±4 |
| 4.03 | 0.04 | 0.04 | 30 | 1.5 | 25.4±3.5 |

**Reference**

[1] S.R. de Groot and P. Mazur, *Non-Equilibrium Thermodynamics* (Dover, 1984).

[2] L.J.T.M. Kempers, Journal of Chemical Physics **115**, 6330 (2001).

[3] A. Firoozabadi, K. Ghorayeb, K. Shukla, AIChE Journal **46**, 892 (2000).

[4] H. Brenner, Physical Review E **74**, 036306 (2006).

[5] J. Rauch, M. Hartung, A.F. Privalov and W. Köhler, Journal of Chemical Physics **126**, 214901 (2007).

[6] M.E. Schimpf and S.N. Semenov, Journal of NonEquilibrium Thermodynamics **32**, 281 (2007).

[7] A. Parola and R. Piazza, Journal of Physics Condensed Matter **17**, S3639 (2005).

[8] A. Würger, Physical Review Letters **98**, 138301 (2007).

[9] J.K.G .Dhont, S. Wiegand, S. Duhr and D. Braun, Langmuir **23**, 1674 (2007).

[10] J.K. Platten, Journal of Applied Mechanics Transactions ASME **73**, 5 (2006).

[11] B. Hafskjold, T. Ikeshoji and S. Ratkje, Molecular Physics **80**, 1389 (1993).

[12] P. Bordat, D. Reith and F. Müller-Plathe, Journal of Chemical Physics **115**, 8978 (2001).

[13] A. Perronace, G. Ciccotti, F. Leroy, A.H. Fuchs and B. Rousseau, Physical Review E **66**, 031201 (2002).

[14] G. Galliero, B. Duguay, J.P. Caltagirone and F. Montel, Fluid Phase Equi. **208**, 171 (2003).

[15] G. Galliero, B. Duguay, J.P. Caltagirone and F. Montel, Philosophical Magazine **83**, 2097 (2003).

[16] J.-M. Simon, D.K. Dysthe, A.H. Fuchs and B. Rousseau, Fluid Phase Equilibria **150**, 151 (1998).

[17] A. Perronace, C. Leppla, F. Leroy, B. Rousseau and S. Wiegand, Journal of Chemical Physics **116**, 3718 (2002).

[18] M. Zhang and F. Müller-Plathe, Journal of Chemical Physics **123**, 1 (2005).

[19] P. Polyakov, M. Zhang, F. Müller-Plathe and S. Wiegand, Journal of Chemical Physics **127**, 014502 (2007).

[20] C. Nieto-Draghi, J.B. Avalos and B. Rousseau, Journal of Chemical Physics **122**, 1 (2005).

[21] J. Xu, S. Kjelstrup, D. Bedeaux and J.-M Simon, Physical Chemistry Chemical Physics **9**, 969 (2007).

[22] F. Bresme, B. Hafskjold and I. Wold, Journal of Physical Chemistry **100**, 1879 (1996).

[23] M. Zhang and F. Müller-Plathe, Journal of Chemical Physics **125**, 124903 (2006).

[24] J. Colombani, G. Galliero, B. Duguay, J.-P. Caltagirone, F. Montel and P.A. Bopp, Physical Chemistry Chemical Physics **4**, 313 (2002).

[25] G. Galliero, J. Colombani, P.A. Bopp, B. Duguay, J.-P. Caltagirone and F. Montel, Physica A **361**, 494 (2006).

[26] S. Yeganegi and E. Pak, Chemical Physics **333**, 69 (2007).

[27] G. Galliero, M. Bugel and B. Duguay, Journal of NonEquilibrium Thermodynamics **32**, 251 (2007).

[28] P.A. Artola and B. Rousseau, Physical Review Letters **98**, 125901 (2007).

[29] F. Müller-Plathe and D. Reith, Computational and Theoretical Polymer Science **9**, 203 (1999).

[30] M. Vladkov and J.-L. Barrat, NanoLetters **6**, 1224 (2006).

[31] S.K. Das, S.U.S. Choi, H.E. Patel, Heat Transfer Engineering **27**, 3 (2006).

[32] X.Q. Wang and A.S. Mujumdar, International Journal of Thermal Sciences **46**, 1 (2007).

[33] M. Kröger, Physics Reports **390**, 453 (2004).

[34] J.-P. Hansen and I.R. MacDonald, *Theory of Simple Liquids* (Academic Press, 1990).

[35] R.B. Bird, W.E. Stewart and E.N. Lightfoot, *Transport Phenomena* (John Wiley & Sons, New York, 2007).

[36] E.L. Cussler, *Diffusion: Mass Transfer in Fluid Systems* (Cambridge University Press, Cambridge, 1997).

[37] C. Albanese, P. Dell'Aversana and F.S. Gaeta, Physical Review Letters **79**, 4151 (1997).

[38] H.J.C Berendsen, J.P.M. Postma, W.F. van Gunsteren, A. di Nola and J.R. Haak, Journal of Chemical Physics **81**, 3684 (1984).

[39] M. Cappelezzo, C.A. Capellari, S.H. Pezzin and L.A.F Coehlo, Journal of Chemical Physics **126**, 224516, 2007.

[40] F. Ould-Kaddour and D. Levesque, Physical Review E **63**, 011205 (2001).

[41] G. Galliero, Fluid Phase Equilibria **224,** 13 (2004).

[42] S. Pan, M.Z. Saghir, M. Kawaji, C.G. Jiang, Y. Yan, Journal of Chemical Physics **126**, 014502 (2007).

[43] H. Ning, R. Kita, H. Kriegs, J. Luettmer-Strathmann and S. Wiegand, Journal of Physical Chemistry B **110**, 10746 (2006).

[44] E. Brinquier and A. Bourdon, Physica A **385**, 9 (2007).

[45] R.B. Bird, C.F. Curtiss, K.J. Beers, Rheologica Acta **36**, 269 (1997).

[46] W. Evans, J. Fish, P. Keblinski, Applied Physics Letters **88** 093116 (2006).

[47] M. Schimpf and J.C. Giddings, Macromolecules **20**, 1561 (1987).

[48] M. Hartung, J. Rauch and W. Köhler, Journal of Chemical Physics **125**, 214904 (2006).

[49] D. Vigolo, G. Brambilla and R. Piazza, Physical Review E **75**, 040401 (2007).

[50]     J.P. Boon and S. Yip, *Molecular Hydrodynamics*, (McGraw-Hill, 1980).

[51]     J. Luettmer-Strathmann, *International Journal of Thermophysics* **26**, 1693 (2005).

[52]     F.S. Gaeta, Physical Review **182**, 289 (1969).

[53]     D.M. Heyes, M.J. Nuevo, J.J. Morales and A.C. Branka, Journal of Physics Condensed Matter **10**, 10159 (1998).

**Table:**

**Table I**: Relation between numbers of atoms inside a nanoparticle and its effective diameter for the various sizes tested.

| $N_p$ | 13 | 55 | 177 | 381 | 767 | 1289 |
|---|---|---|---|---|---|---|
| $d_{NP}/\sigma$ | 2.4 | 4.03 | 5.98 | 7.73 | 9.75 | 11.59 |

**Figure captions:**

**Figure 1**: Stationary temperature and nanoparticles molar fraction profiles in a half simulation box ($d_{NP}/\sigma$=4.03 and nanoparticles volume fraction equal to 7.31 %)

**Figure 2**: Thermal diffusion in "isotopic" LJ mixtures ($m_1/m_2$=10, $\sigma_1=\sigma_2$ and $\varepsilon_1=\varepsilon_2$) for two states, circles correspond to $T^*$=1 and $\rho^*$=0.8, squares correspond to $T^*$=1.686 and $\rho^*$=0.477. Open symbols have been obtained using the usual NEMD approach and full symbols through the SPTA. Dotted lines correspond to the correlation proposed in Ref. 27.

**Figure 3**: Thermal diffusion factors versus nanoparticles volume fraction for two nanoparticle sizes, $d_{NP}/\sigma$=2.4, circles, $d_{NP}/\sigma$=4.03, squares. Open symbols correspond to usual NEMD results, full symbols to SPTA ones.

**Figure 4**: Thermal conductivity in nanofluids for two nanoparticles sizes, $d_{NP}/\sigma$=2.4, circles, $d_{NP}/\sigma$=4.03, squares. Full up triangle corresponds to the pure fluid value.

**Figure 5**: Single particle thermal diffusion coefficient and mutual diffusion versus the inverse of the diameter of the nanoparticle. Circles correspond to $D_T^{SP}$ and squares to $D_{12}^{SP}$.

**Figure 6**: Influence of the quality of the solvent on $D_T^{SP}$, circles, and $D_{12}^{SP}$, squares.

**Figure 7**: Influence of the viscosity of the solvent on $D_T^{SP}$, circles, and $D_{12}^{SP}$, squares.

**Figure 8**: Influence of the mass of the atoms constituting the nanoparticle on $D_T^{SP}$, circles, and $D_{12}^{SP}$, squares.

**Figure 9**: Influence of the spring constant of the FENE potential, $\kappa$, on $D_T^{SP}$, circles, and $D_{12}^{SP}$, squares.

**Figure 1**

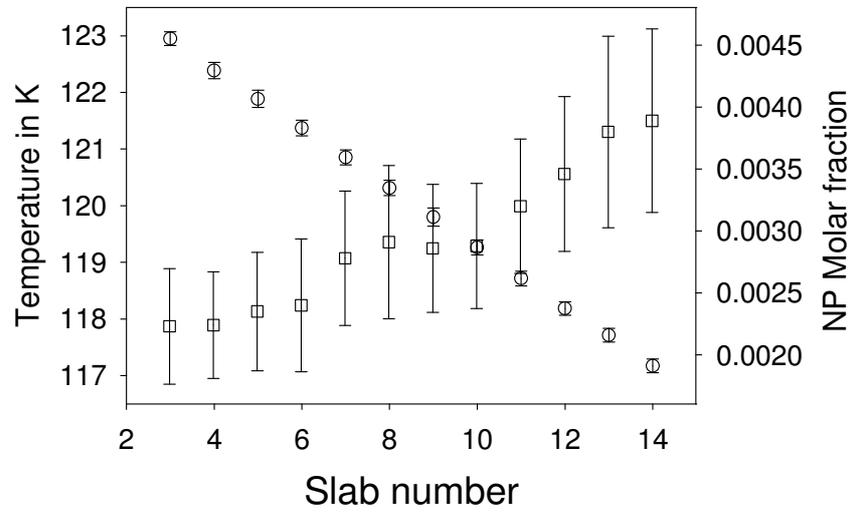

**Figure 2**

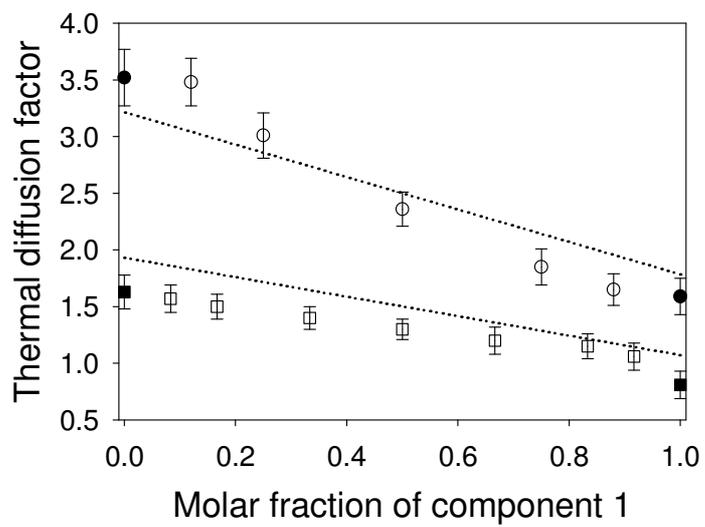

**Figure 3**

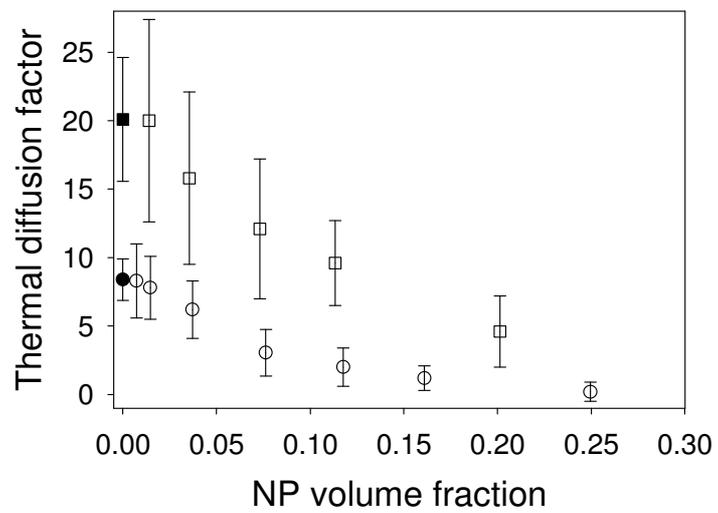

**Figure 4**

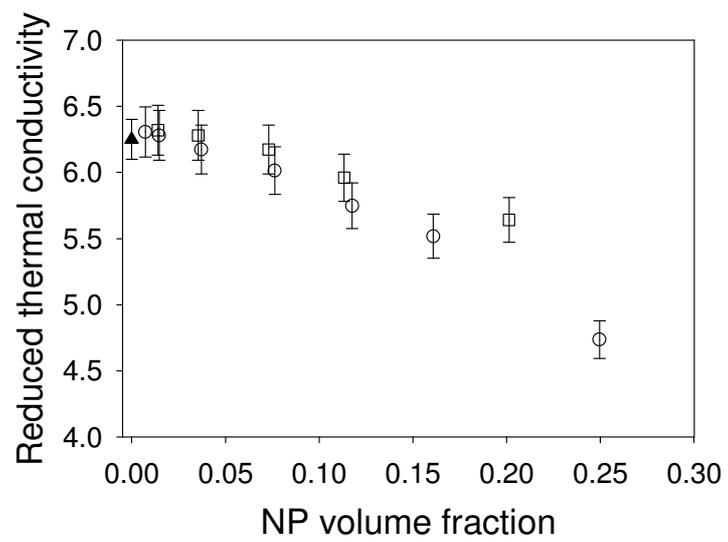

**Figure 5**

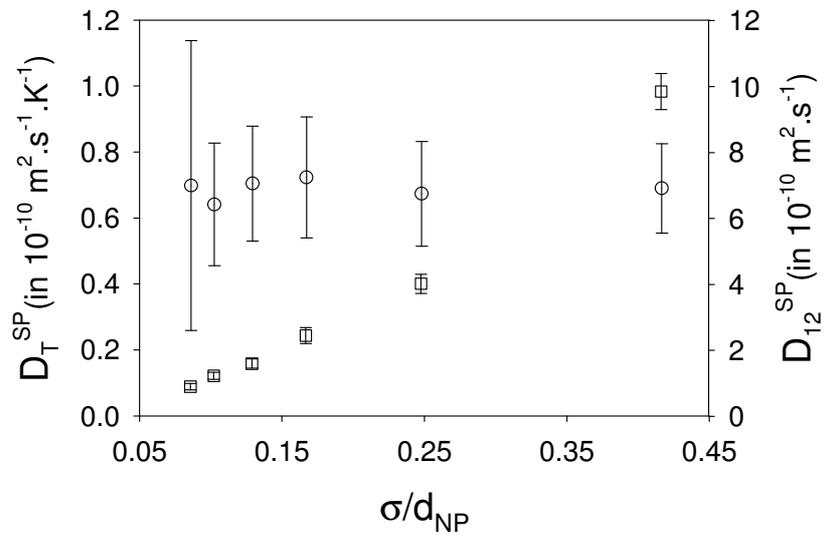

**Figure 6**

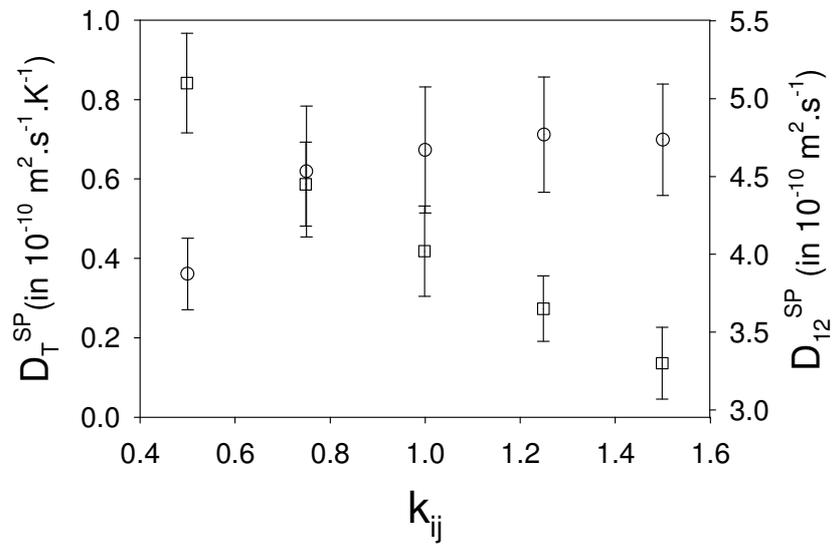

**Figure 7**

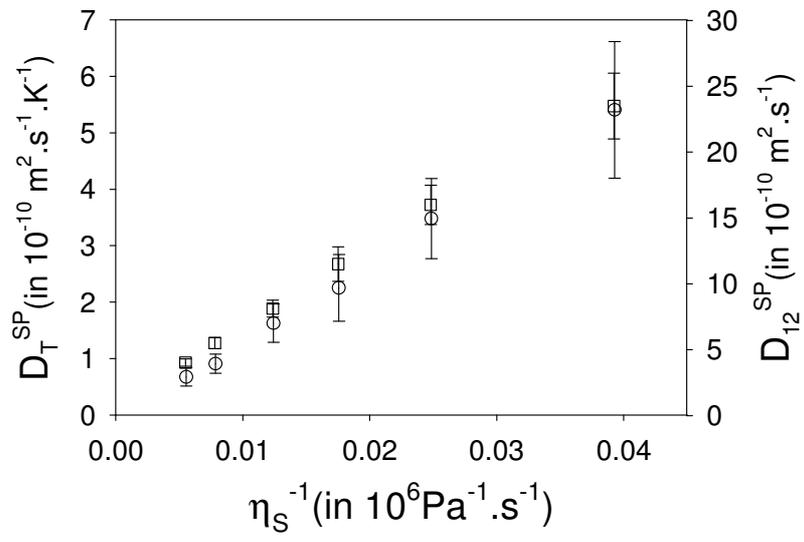

**Figure 8**

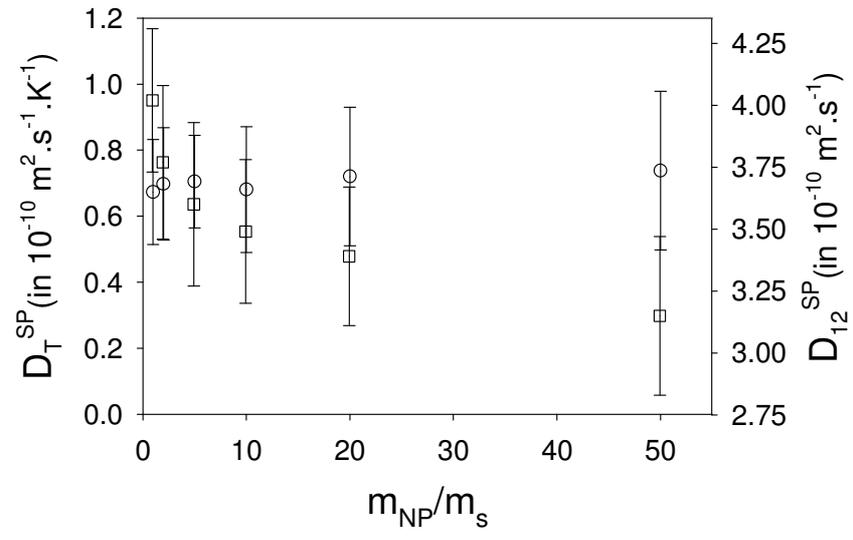

**Figure 9**

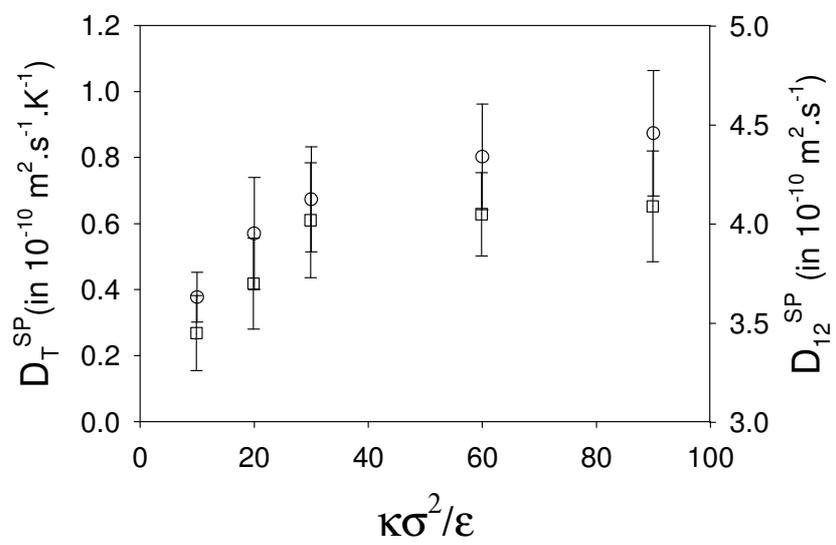